\documentclass[useAMS,usenatbib,usegraphicx]{mn2e}
\usepackage{color}
\usepackage{wasysym}
\usepackage{url}
\usepackage{multirow}

\newcommand{\kms}{\ensuremath{\mathrm{km\, s^{-1}}}}

\newcommand{\mm}{\ensuremath{\mathrm{mm}}}

\newcommand{\nuc}[2]{\ensuremath{\mathrm{^{#1}#2}}}
\newcommand{\ions}[2]{#1\,{\sc #2}}
\newcommand{\ye}{\ensuremath{Y_\mathrm{e}}}
\newcommand{\msun}{\ensuremath{\mathrm{M}_\odot}}
\newcommand{\gccm}{\ensuremath{\mathrm{g} \, \mathrm{cm}^{-3}}}

\def\lesssim{\mathrel{\hbox{\rlap{\hbox{\lower4pt\hbox{$\sim$}}}\hbox{$<$}}}}

\def\gtrsim{\mathrel{\hbox{\rlap{\hbox{\lower4pt\hbox{$\sim$}}}\hbox{$>$}}}}

\def\aj{AJ}%
\def\araa{ARA\&A}%
\def\apj{ApJ}%
\def\apjl{ApJ}%
\def\apjs{ApJS}%
\def\aap{A\&A}%
\def\mnras{MNRAS}%
\def\pasp{PASP}%
\def\nat{Nature}%
\title[2002cx-like SNe from deflagrations with bound remnants]
{%
  3D deflagration simulations leaving bound remnants:
  a model for 2002cx-like Type Ia supernovae%
  \thanks{Based on observations collected at ESO,
    Paranal. Program ID 076.D-0183}
}

\author[M.~Kromer~et~al.]
{M.~Kromer,$^{1}$\thanks{E-mail: mkromer@mpa-garching.mpg.de}
 M.~Fink,$^{2}$ 
 V.~Stanishev,$^{3}$
 S.~Taubenberger,$^{1}$
 \newauthor
 F.~Ciaraldi-Schoolman,$^{1}$
 R.~Pakmor,$^{4}$
 F.~K.~R\"{o}pke,$^{2}$
 A.~J.~Ruiter,$^{1}$
 \newauthor
 I.~R.~Seitenzahl,$^{2}$
 S.~A.~Sim,$^{5}$
 G.~Blanc,$^{6,7}$
 N.~Elias-Rosa$^{8}$
 and W.~Hillebrandt$^{1}$\\
 $^{1}$Max-Planck-Institut f{\"u}r Astrophysik, 
       Karl-Schwarzschild-Str. 1, D-85748 Garching bei M{\"u}nchen, Germany\\
 $^{2}$Institut f\"ur Theoretische Physik und Astrophysik, Universit\"at
       W\"urzburg, Emil-Fischer-Stra{\ss}e 31, D-97074 W\"urzburg, Germany\\
 $^{3}$CENTRA - Centro Multidisciplinar de Astrof\'isica, Instituto 
      Superior T\'ecnico, Av. Rovisco Pais 1, 1049-001 Lisbon, Portugal\\
 $^{4}$Heidelberger Institut f\"{u}r Theoretische Studien, 
       Schloss-Wolfsbrunnenweg 35, D-69118 Heidelberg, Germany\\
 $^{5}$Research School of Astronomy \& Astrophysics, 
       Mount Stromlo Observatory, Cotter Road, Weston ACT 2611, Australia\\
 $^{6}$INAF Osservatorio Astronomico di Padova, 
       Vicolo dell'Osservatorio 5, 35122 Padova, Italy\\
 $^{7}$Universit\'e Paris Diderot-Paris 7,
       Laboratoire APC, 10 rue Alice Domon et L\'eonie Duquet, 75205 Paris cedex 13, France\\
 $^{8}$Institut de Cincies de l'Espai (IEEC-CSIC), 
       Facultat de Cincies, Campus UAB, 08193 Bellaterra, Spain
}

\begin{document}

\date{Accepted 2012 November 26}

\pagerange{\pageref{firstpage}--\pageref{lastpage}} \pubyear{2012}

\maketitle

\label{firstpage}

\begin{abstract}
  2002cx-like supernovae are a sub-class of sub-luminous Type~Ia 
  supernovae.  Their light curves and spectra are characterized by 
  distinct features that indicate strong mixing of the explosion
  ejecta.  Pure turbulent deflagrations have been shown to produce 
  such mixed ejecta.  Here, we present hydrodynamics, nucleosynthesis
  and radiative transfer calculations for a 3D full-star deflagration
  of a Chandrasekhar-mass white dwarf.  Our model is able to reproduce 
  the characteristic observational features of SN~2005hk (a 
  proto-typical 2002cx-like supernova), not only in the optical,
  but also in the near-infrared.  For that purpose we present,
  for the first time, five near-infrared spectra of SN~2005hk 
  from $-0.2$ to $26.6$ days with respect to $B$-band maximum.  
  Since our model burns
  only small parts of the initial white dwarf, it fails to completely 
  unbind the white dwarf and leaves behind a bound remnant of 
  $\sim$\,1.03\,\msun\ -- consisting mainly of unburned carbon and 
  oxygen, but also enriched by some amount of intermediate-mass and 
  iron-group elements from the explosion products that fall back on 
  the remnant.  We discuss possibilities for detecting this bound 
  remnant and how it might influence the late-time observables of 
  2002cx-like SNe.
\end{abstract}


\section{Introduction}

Due to an empirical relation between their peak brightness and their 
light curve evolution \citep{phillips1993a,pskovskii1977a}, Type Ia 
supernovae (SNe~Ia) can be used as standardiseable candles to measure 
the expansion history of the Universe \citep{riess1998a, schmidt1998a, 
perlmutter1999a}.  However, we still
have no detailed picture about the progenitor systems of these luminous
explosions beyond the general idea that they originate from the
thermonuclear disruption of carbon--oxygen (CO) white dwarfs (WDs)
\citep[for a review see e.g.][]{hillebrandt2000a}.  Radioactive 
isotopes, in particular \nuc{56}{Ni} and its daughter nucleus 
\nuc{56}{Co}, which are synthesised within the thermonuclear burning, 
power the observable display of SNe~Ia.

As of today several sub-classes of SNe~Ia have been identified 
\citep{li2011a}.  Among the most peculiar of these events are the 
SN~2002cx-like objects \citep{li2003a}.  For their light curve
decline rate, these objects are sub-luminous with respect to the
Phillips relation and do not show secondary maxima in the near-infrared
(NIR) bands. 
Their spectra are characterised by very low expansion velocities 
compared to other SNe~Ia and show signs of strong mixing in the 
ejecta \citep{jha2006a,phillips2007a}.  This is in clear contrast to
normal SNe~Ia, which are characterised by strongly layered ejecta 
\citep{stehle2005a,mazzali2007a}.

While explosion models involving a detonation are not able to
explain such an ejecta structure 
\citep[e.g.][]{sim2010a,seitenzahl2012a}, such a strong 
mixing can easily be obtained from turbulent deflagrations in 
WDs (\citealt*{gamezo2004a}; \citealt{roepke2005b}).  The low expansion
velocities are also in good agreement with the small
amount of kinetic energy released in deflagration models. Such 
considerations led
\citet{branch2004a} and \citet{jha2006a} to conclude that 
SN~2002cx-like objects might be related to pure deflagrations of
Chandrasekhar-mass WDs.  This interpretation was supported 
by \citet{phillips2007a}, who compared the broad-band light curves
of SN~2005hk (a well-sampled proto-typical 02cx-like event) 
to synthetic light curves of a 3D deflagration model of 
\citet{blinnikov2006a} and found good agreement between the data 
and the model.

The models presented by \citet{blinnikov2006a}, however, do not
allow for a detailed comparison to the observed spectral time sequence
since the multi-group approximation employed in their radiative 
transfer simulations is too coarse.  Moreover, their underlying 
hydrodynamic explosion models are restricted to one spatial octant
of the progenitor WD, introducing artificial symmetries to the flame
evolution.  Here, we report on a 3D full-star deflagration simulation
through to the homologous expansion phase.  We perform 
detailed radiative transfer calculations with the time-dependent 
3D radiative transfer code {\sc artis} \citep{kromer2009a,sim2007b} 
from which we obtain a time series of synthetic spectra that we 
compare to the observed spectra of SN~2005hk.

The paper is organised as follows.  In Section~\ref{sec:explosion}
we give a brief description of our explosion simulation and present
the resulting ejecta structure.  In Section~\ref{sec:observables} 
we present synthetic observables for this explosion model and
compare them to the observed light curves and spectra of SN~2005hk.
Finally, we discuss our results and give conclusions in 
Sections~\ref{sec:discussion} and \ref{sec:conclusion}, respectively.

\section{Explosion simulation}
\label{sec:explosion}

A Chandrasekhar-mass WD is believed to undergo about a century of
convective carbon burning in the centre before a thermonuclear
runaway finally sets in.  Since this so-called simmering phase is 
characterised by highly turbulent flows, it cannot be fully accounted
for by present-day numerical simulations (but see e.g.\ \citealt{hoeflich2002a}; \citealt*{kuhlen2006a}; \citealt{zingale2009a,nonaka2012a}). Thus, the 
actual ignition configuration of Chandrasekhar-mass WDs is not 
well constrained.  Given this ignorance, one may use the ignition 
geometry as a free parameter and explore a larger set of explosion
simulations with various ignition setups.  Thereby one has to account
for both different ignition strengths and ignition positions.  A
good way to achieve the former is to use a multi-spot ignition 
scheme which seeds unstable burning modes in a robust and 
numerically well-controlled way.  Recently, we have performed such a 
systematic study for different ignition setups of 3D full-star pure 
deflagration simulations (Fink et al., in preparation) yielding 
\nuc{56}{Ni} masses between 0.035 and 0.38\,\msun.

Here, we focus on a detailed comparison to SN~2005hk. For that
purpose we select one of the models of the series by Fink et al.\
that produces a \nuc{56}{Ni} mass on the order of the observationally
derived value of SN~2005hk.  Applying Arnett's law \citep{arnett1982a}
to the observed bolometric light curve, \citet{phillips2007a} 
report a \nuc{56}{Ni} mass of $\sim$\,0.2\,\msun\ for this SN\@.
With a \nuc{56}{Ni} mass of 0.18\,\msun, model N5def of the Fink 
et al.\ series comes close to that value. 

\begin{figure}
  \centering
  \includegraphics[width=0.9\linewidth]{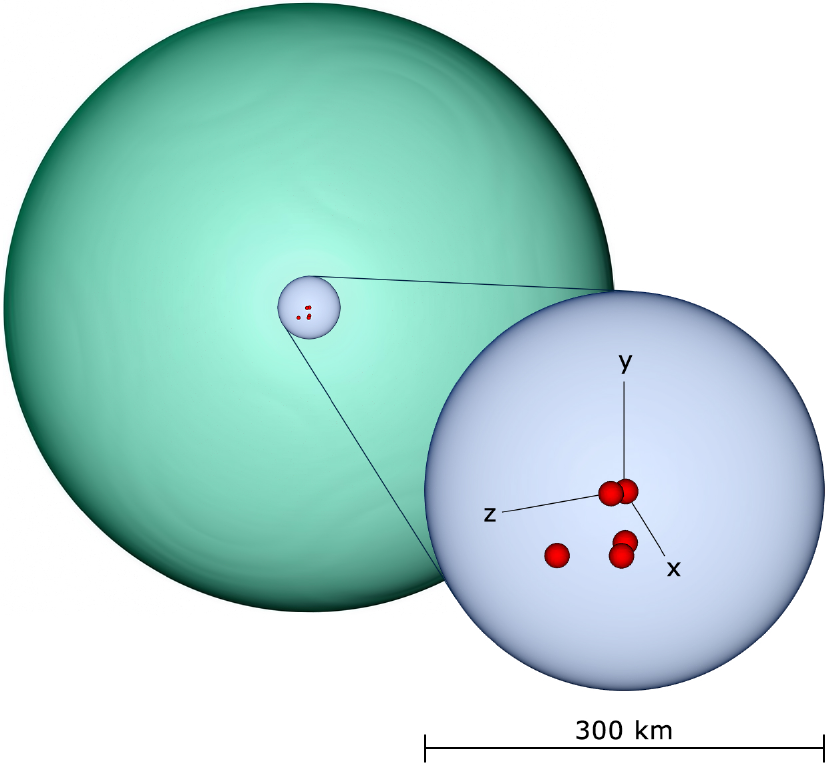}
  \caption{Ignition setup of the N5def model.  Shown is a volume
    rendering of the WD in mint green colour.  As discussed in
    Section~\ref{sec:explosion}, the ignition kernels were randomly 
    placed within a radius of $150\,\mathrm{km}$ around the centre of  
    the WD, as illustrated by the bluish sphere.  The exact 
    configuration of the ignition kernels is shown in the enlarged 
    inlay (see also Table~\ref{tab:ignition}).}
  \label{fig:ignition}
\end{figure}

In the N5def simulation, an isothermal 
($T=5\times10^5\,\mathrm{K}$) Chandrasekhar-mass WD 
was set up in hydrostatic equilibrium with a central density of 
$2.9\times10^9\,\gccm$ and a homogeneous composition of carbon and 
oxygen in equal parts by mass\footnote{Although the 
detailed flame evolution depends on the exact value of the 
central density (which is not well constrained for Chandrasekhar-mass
WDs, e.g.\ \citealt*{seitenzahl2011a}), it is not expected to have a
qualitative impact on the outcome of the explosion.}.  
To account for an assumed solar 
metallicity of the zero-age main-sequence progenitor, we start with 
a $\ye$ of 0.49886, corresponding to 2.5 per cent of \nuc{22}{Ne} 
in the initial composition.
The WD was then discretized
on a three-dimensional Cartesian moving grid \citep{roepke2005c} 
with $512^3$ cells consisting of two nested parts 
(central resolution 1.9\,km) and ignited
in five spherical ignition kernels that were placed randomly in a 
Gaussian distribution within a radius of $150\,\mathrm{km}$ from 
the WD's centre.  By chance, for model N5def this algorithm produced
a fairly one-sided ignition configuration with all kernels 
lying in a relatively small solid angle.  Thus, this configuration is 
representative for a slightly off-centred single-spot ignition with 
a larger number of initially excited burning modes.  The actual 
configuration is shown in Fig.~\ref{fig:ignition}.  All kernels have 
a radius of $r_\mathrm{k} = 10$\,km and are at distances $d$ between 
6.3 and 80.9\,km from the origin (see Table~\ref{tab:ignition}).

\begin{table}
  \caption{Position of the ignition kernels of model N5def.  Given
    are the $x$, $y$ and $z$ coordinates of the centre of the
    individual ignition kernels and their distance $d$ to the 
    centre of the WD.}
  \label{tab:ignition}
  \centering
  \begin{tabular}{lrrrr}
    \hline
    \#  & $x$    & $y$     & $z$    & $d$ \\
         & \multicolumn{4}{c}{(in km)} \\
    \hline
    1    & $65.5$ & $-15.5$ & $24.0$ & $71.5$ \\
    2    & $38.6$ & $-22.7$ & $67.3$ & $80.9$ \\
    3    & $13.0$ & $  8.2$ & $15.1$ & $21.6$ \\
    4    & $-5.0$ & $-51.3$ & $-2.6$ & $51.7$ \\
    5    & $ 5.6$ & $  2.9$ & $ 0.6$ & $6.3$  \\
    \hline
  \end{tabular}
\end{table}

\begin{figure*}
  \centering
  \includegraphics{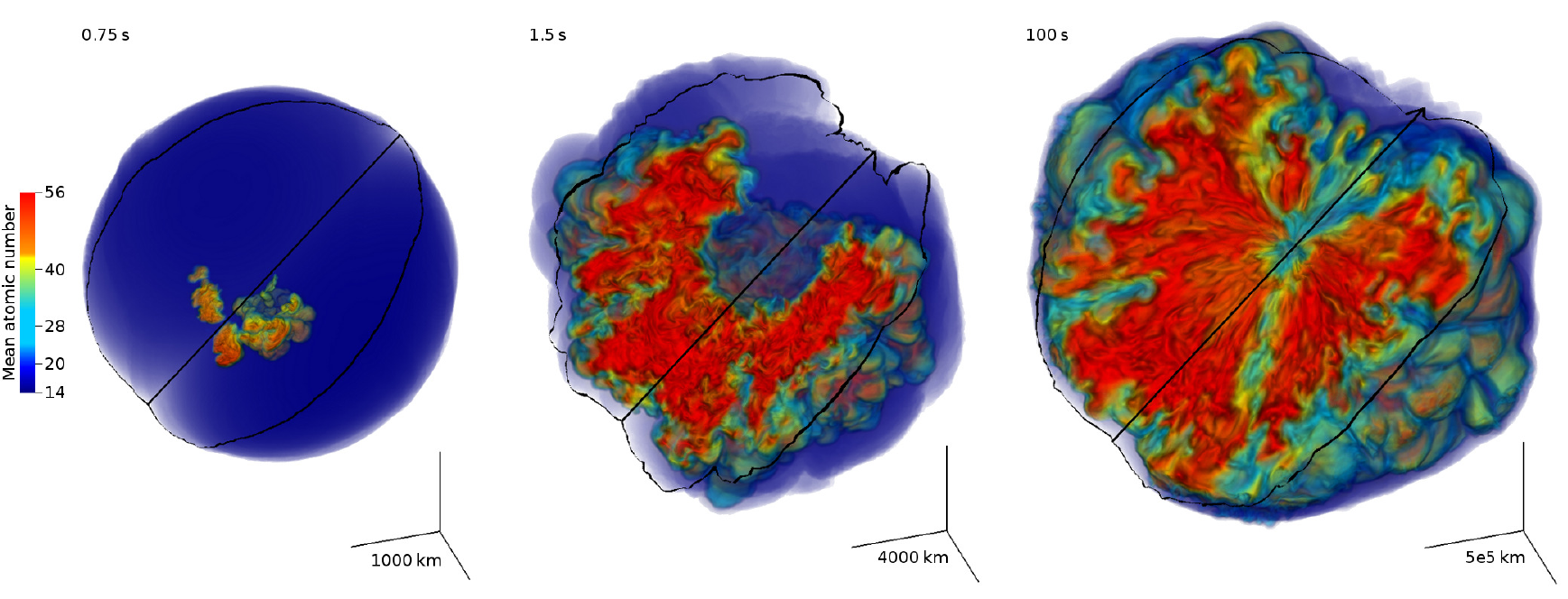}
  \caption{Snapshots of the hydrodynamic evolution of our model N5def.
    Shown are volume renderings of the mean atomic number calculated
    from the reduced set of species in the hydrodynamic simulation
    (see colour
    bar).  To allow a view to the central part of the ejecta, a wedge
    was carved out from the front of the ejecta.  (i) At 0.75\,s a 
    one-sided deflagration plume rises towards the WD surface and
    fragments due to Rayleigh-Taylor and Kelvin-Helmholtz 
    instabilities.  (ii) At 1.5\,s the expansion of the WD quenches
    the burning and the explosion ashes wrap around the unburned core.  
    (iii) Finally, at 100\,s the 
    unburned core is completely engulfed by the explosion ashes 
    which are ejected into space.  The small triads at the bottom
    right corner of each panel indicate the scaling at the origin
    of each plot: from left to right the legs of the triads represent 
    1000\,km, 4000\,km and 500\,000\,km, respectively.}
  \label{fig:hydro}
\end{figure*}

Neglecting any possible deflagration-to-detonation transition 
(see \citealt{seitenzahl2012a} for an alternative evolution 
of this model in a delayed-detonation scenario) we followed the flame
evolution up to 100\,s after ignition when the ejecta reach 
homologous expansion. To this end we used the LEAFS code, a 
three-dimensional finite-volume discretization of the reactive Euler 
equations which is based on the PROMETHEUS implementation
\citep*{fryxell1989a} of the `piecewise parabolic method' (PPM) by
\citet{colella1984a}. Deflagration fronts are modelled as 
discontinuities between carbon--oxygen fuel and nuclear ash, and their
propagation is tracked with a level-set scheme (\citealt{reinecke1999a,
osher1988a}; \citealt*{smiljanovski1997a}).  All material crossed by these 
fronts is converted to nuclear ash with a composition and energy release
depending on fuel density.  Composition and energy release are
interpolated from tables which have been calibrated using our full
384-isotope network.  After a very short phase of laminar burning
following ignition, the propagation of deflagrations is dominated by
buoyancy- and shear-induced instabilities and interactions with a
complex turbulent flow field.  The unresolved acceleration of the flame
due to turbulence is accounted for by a sub-grid scale model
(\citealt*{schmidt2006b}; \citealt{schmidt2006c}).  Self-gravity is 
dealt with by a monopole gravity solver.

The hydrodynamic evolution of our model is shown in 
Fig.~\ref{fig:hydro}.  Since a deflagration flame cannot burn 
against the density gradient, our asymmetric ignition configuration 
leads to the formation of a one-sided deflagration plume that
fragments due to Rayleigh-Taylor and Kelvin-Helmholtz instabilities.
Once the deflagration front reaches the outer layers of the WD, the
burning quenches due to the expansion of the WD and the ashes wrap 
around the still unburned core until they finally engulf it
completely.  A similar evolution of the deflagration flame was already 
described for single-spot off-centre ignitions by e.g.\ \citet*{plewa2004a} 
and \citet*{roepke2007a}.  While \citet{plewa2004a} found an ensuing
detonation to be triggered when the ashes collide on the far side
of the star (see also \citealt{seitenzahl2009c}), we -- similarly 
to \citet{roepke2007a} and \citet{jordan2012b} -- do not find high 
enough densities and temperatures for such a detonation to occur due 
to a significant expansion of the WD during the deflagration phase.

Since only a moderate fraction of the core of the WD is burned,
the nuclear energy release 
in our simulation ($E_\mathrm{nuc}=4.9\times 10^{50}$\,erg) is less 
than the binding energy of the initial WD ($E_\mathrm{bind}=
5.2\times 10^{50}$\,erg).  Nevertheless, about 0.37\,\msun\ of the WD 
are accelerated to escape velocity and ejected into the ambient 
medium with a kinetic energy of $1.34\times10^{50}$\,erg.  The 
remainder of the mass of the initial WD is left behind and forms a
{\em bound remnant}.  Similar findings were already reported in 
2D \citep*[e.g.][]{livne2005a} and recently in the context of a failed 
gravitationally-confined detonation (\citealt{jordan2012b}).  

To obtain detailed nucleosynthesis yields of the explosion, we
performed a post-processing calculation with a 384 isotope network for
$10^6$ Lagrangian tracer particles which were passively advected during 
the hydrodynamic simulation to record thermodynamic trajectories of
mass elements \citep{travaglio2004a, seitenzahl2010a}.  A compilation
of the masses of important species is given in Table~\ref{tab:yields}.
To determine the mass that stays bound in the remnant and that of the
unbound ejecta, we calculated the asymptotic specific kinetic energy
$\epsilon_\mathrm{kin,a} = \epsilon_\mathrm{kin,f} +
\epsilon_\mathrm{grav,f}$ for all tracer particles.  Here,
$\epsilon_\mathrm{kin,f} = v_\mathrm{f}^2 / 2$ and
$\epsilon_\mathrm{grav,f}$ are the specific kinetic and gravitational
binding energies at $t = 100\,\mathrm{s}$, i.e.\ at the end of our
simulation, respectively.  Only for positive values of
$\epsilon_\mathrm{kin,a}$ a particle will be able to escape the
gravitational potential.  Otherwise, it will stay bound.

\begin{table}
  \caption{Yields of select species for model N5def.}
  \label{tab:yields}
  \centering
  \begin{tabular}{lcc}
    \hline
     & Bound remnant & Ejecta\\
     & (\msun) & (\msun) \\
    \hline
    Total        & 1.028  & 0.372  \\
    C            & 0.422  & 0.043  \\
    O            & 0.484  & 0.060  \\
    Ne           & 0.054  & 0.005  \\
    Mg           & 0.004  & 0.013  \\
    Si           & 0.015  & 0.025  \\
    S            & 0.004  & 0.009  \\
    Ca           & 0.0003 & 0.001  \\
    Fe           & 0.004  & 0.031  \\
    Ni           & 0.025  & 0.187  \\
    \nuc{56}{Ni} & 0.022  & 0.158  \\
    \hline
  \end{tabular}
\end{table}

\begin{figure*}
  \centering
  \includegraphics[width=0.8\linewidth]{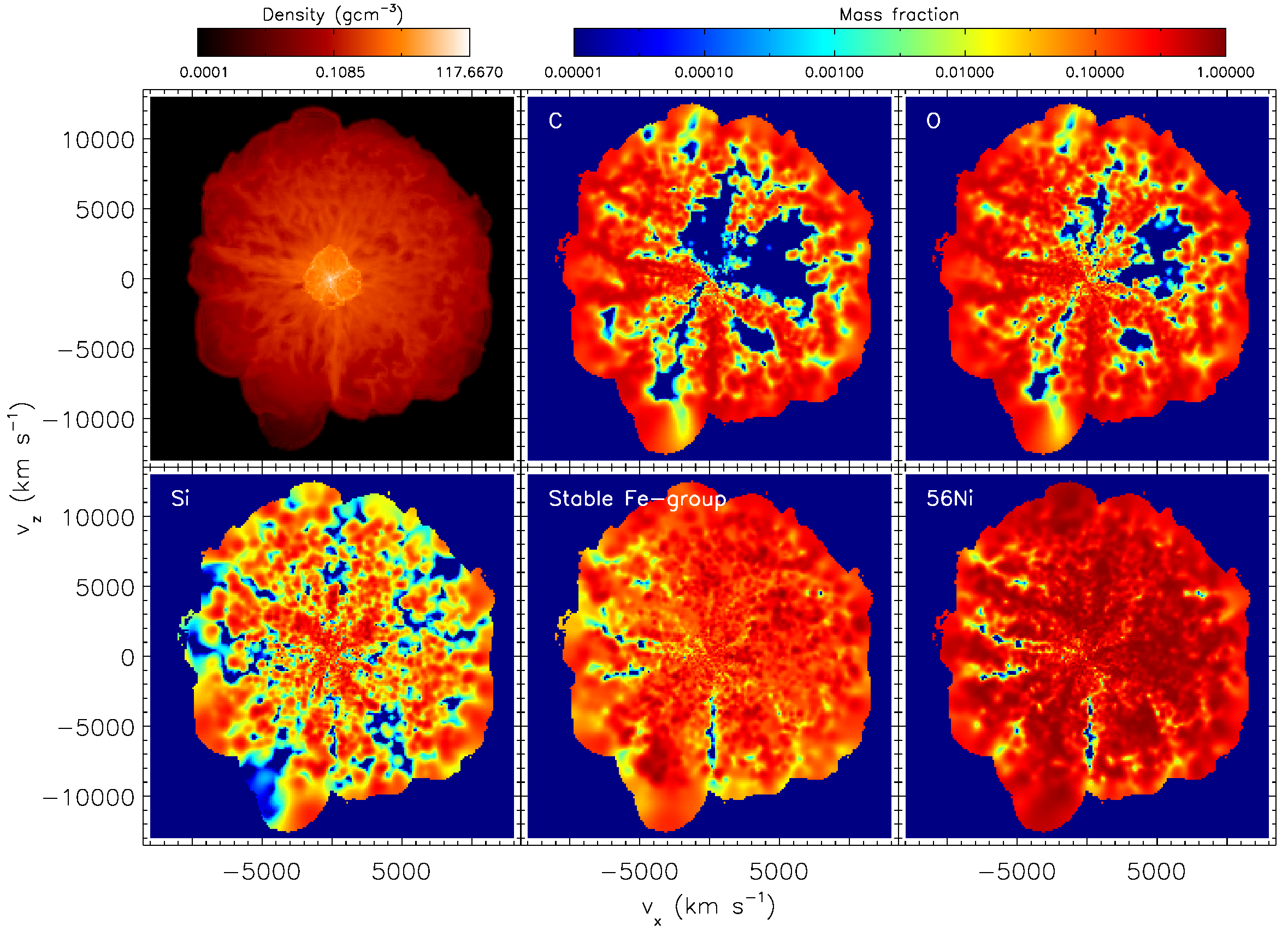}
  \caption{Composition of the unbound ejecta of model N5def. 
    Shown are slices through the $x$--$z$-plane in the asymptotic 
    velocity space at 100\,s after the explosion. The top left 
    panel displays the density, the following panels show the mass 
    fraction of selected species (C, O, Si, stable iron-group 
    elements and \nuc{56}{Ni} from top left to bottom right).
    }
  \label{fig:composition}
\end{figure*}

Finally, we mapped the unbound tracer particles to a Cartesian
grid to reconstruct the chemical composition of the explosion
ejecta in the asymptotic velocity space.\footnote{For unbound tracer 
particles, the asymptotic velocity is determined as $v_\mathrm{a} = 
\sqrt{2 \epsilon_\mathrm{kin,a}}$.}  To this end, we used an 
SPH like algorithm as described
in \citet{kromer2010a}.  The resulting ejecta structure is shown
in Fig.~\ref{fig:composition}.  In contrast to the simulation
by \citet{jordan2012b}, our model shows no pronounced large-scale
composition asymmetry.  There are, however, small scale anisotropies
due to the turbulent evolution of the deflagration flame. 
The density jump at about 2000\,\kms\ is produced during the strongest
pulsation of the bound core (at $t \sim 12$\,s): After significant
initial expansion due to burning, the mostly unburned inner parts of
the ejecta start falling back inwards.  This leads to the formation 
of a central compact core, which, after maximum compression, starts to 
expand again.  At the same time, outer layers are still infalling and 
an accretion shock forms at the edge of the dense core 
\citep[cf.][]{bravo2009a}.  In our model, maximum temperatures in the 
accretion shock just reach $10^9$\,K at densities of $5 \times 10^5\,
\gccm$, therefore no detonation will be triggered (for critical 
conditions see \citealt{roepke2007a} and \citealt{seitenzahl2009b}).

\section{Synthetic observables}
\label{sec:observables}

To obtain synthetic spectra and light curves for our model, we used 
the time-dependent 3D Monte Carlo radiative transfer code 
\textsc{artis} \citep{kromer2009a,sim2007b}.  For the radiative
transfer simulation we mapped the abundance and density structure
of the unbound ejecta at the end of the hydrodynamic simulations (by 
which point homologous expansion is a good approximation)
to a $50^3$ Cartesian grid and followed the propagation of $10^8$ 
photon packets for 111 logarithmically-spaced time steps between 2 
and 120\,d after explosion.  To speed up the initial phase, a grey
approximation, as discussed by \citet{kromer2009a}, was used in
optically thick cells, and the initial 10 time steps (i.e.\ the initial
three days after the explosion) were treated in local thermodynamic 
equilibrium (LTE).  For our simulation we used the `big\url{_}gf-4' 
atomic data set of \citet{kromer2009a} with a total of 
${\sim}\,8.2\times10^6$ bound--bound transitions.
The resulting broad-band light curves and a spectral time series
are shown in Figs.~\ref{fig:lightcurves} and \ref{fig:specplot} 
and compared to SN~2005hk as a proxy for SN~2002cx-like objects.

\subsection{Broad-band light curves}
\label{sec:lightcurves}

\begin{figure*}
    \centering
  \includegraphics{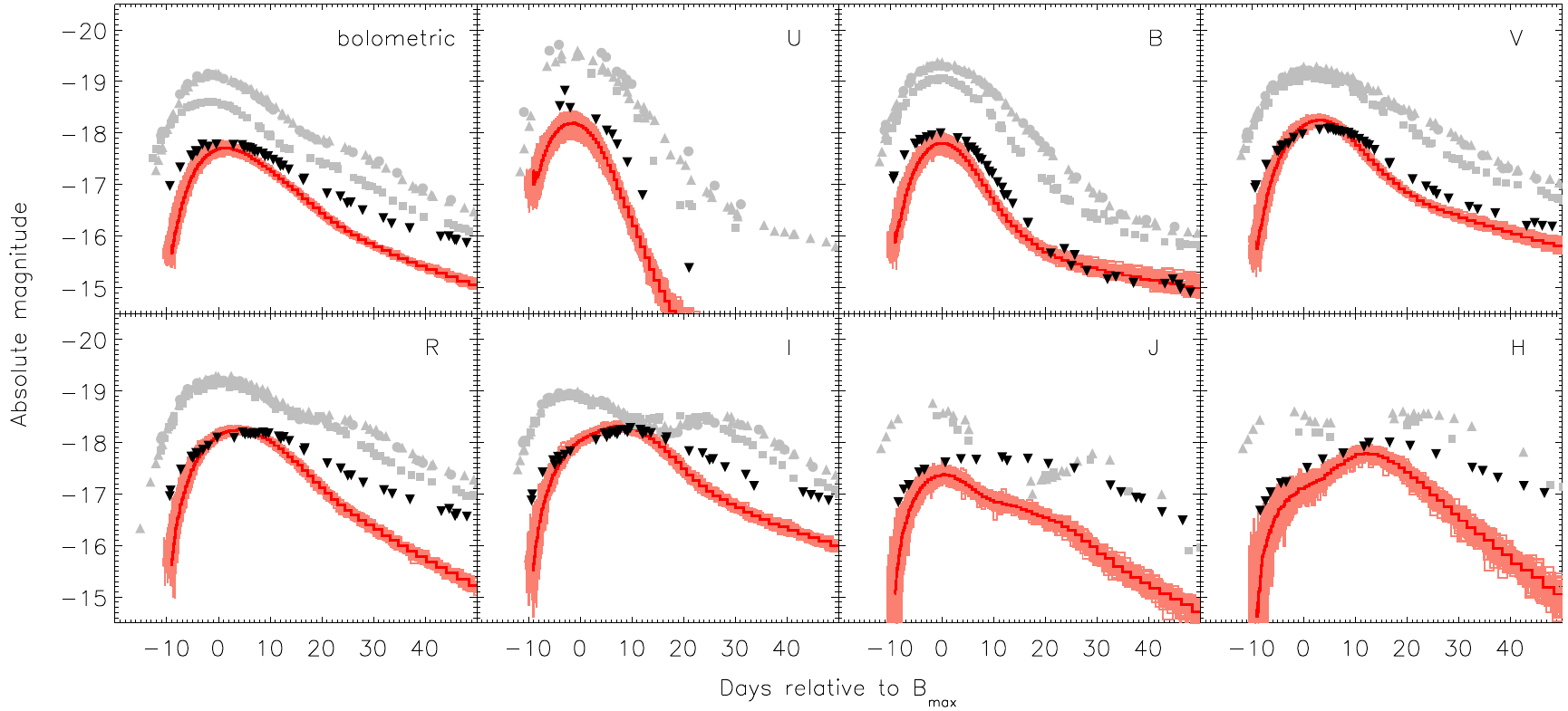}
  \caption{Synthetic light curves of our model.  The panels from 
    top left to bottom right contain \emph{UBVRIJHK} bolometric and 
    broad-band $U$,$B$,$V$,$R$,$I$,$J$,$H$ light curves
    \citep{bessell1988a, bessell1990a}. 
    The dark red line corresponds to the angle-average of the model. 
    The light red region indicates the scatter caused by 100 different 
    viewing angles.  Time is given relative to $B$-band maximum.  The 
    black upside-down triangles show the observed light curves of 
    SN~2005hk \citep{phillips2007a}.  Due to missing $K$-band data
    for SN~2005hk the bolometric light curve shown only accounts for
    \emph{UBVRIJH} fluxes.  For comparison, the grey symbols 
    show observational data of three well-observed normal SNe~Ia: 
    SN~2003du \citep[circles, ][]{stanishev2007b}, SN~2004eo 
    \citep[squares, ][]{pastorello2007b}, and SN~2005cf 
    \citep[triangles, ][]{pastorello2007a}.}
    \label{fig:lightcurves}
\end{figure*}

As reported by \citet{jha2006a} and \citet{phillips2007a},
SN~2002cx-like objects are characterised by distinct spectral 
features and light curves. In the following, we will investigate
to which extent our model is capable to reproduce these features.
From the broad-band light curves in Fig.~\ref{fig:lightcurves} 
it is immediately obvious that our model matches the \emph{low 
peak luminosities} of 2002cx-like SNe quite well. This, of course,
is not so surprising, since we picked a model with a \nuc{56}{Ni} mass
close to that observationally derived for SN~2005hk. However,
we do not only get the correct peak luminosities but also approximately
the right colours at maximum. Moreover, our model naturally 
explains the \emph{absence of secondary maxima in the NIR} bands.
This is a consequence of the turbulent burning, which leads to
an almost completely mixed ejecta structure with a roughly
constant iron-group element (IGE) mass fraction at all 
velocities (see Figs.~\ref{fig:hydro} and \ref{fig:composition}).  
\citet{kasen2006b} has shown that SN ejecta with such a homogenized 
composition are not expected to show secondary maxima in the NIR.

Other peculiarities of 2002cx-like SNe are their \emph{slowly 
declining light curves in $R$ and redder bands} and consequently 
also in UVOIR bolometric.  In this respect our model has some 
shortcomings.  While it nicely reproduces the post-maximum decline 
in the $U$, $B$ and $V$ bands, our post-maximum light curve evolution 
is too fast particularly in $R$ and redder bands, but also in UVOIR
bolometric.  This could be a consequence of our low ejecta mass of 
only 0.37\,\msun, which is not able to trap enough $\gamma$-radiation 
after maximum.  We note, however, that the synthetic light curves 
in Fig.~\ref{fig:lightcurves} are only powered by radionuclides in
the ejecta.  Thereby we neglect any possible influence from the bound 
remnant, a puffed-up stellar object heated by the explosion, and 
also from the \nuc{56}{Ni} that falls back onto the remnant 
(see Table~\ref{tab:yields}).  After maximum light, when the ejecta 
start to become optically thin, the bound remnant will be exposed 
and it could kick in as an additional luminosity source. We will
discuss this intriguing possibility in Section~\ref{sec:remnant}.

However, looking at the overall shape of the early light curves,
an insufficient trapping of radiation in the ejecta and thus 
too low ejecta mass seems to be more likely.  With a $B$-band 
rise time of 11.2\,d, our model evolves somewhat too fast compared 
to the rise time of SN~2005hk, for which \citep{phillips2007a} find 
a value of $15\pm1$\,d.

\subsection{Optical spectra}

\begin{figure}
  \centering
  \includegraphics[width=\linewidth]{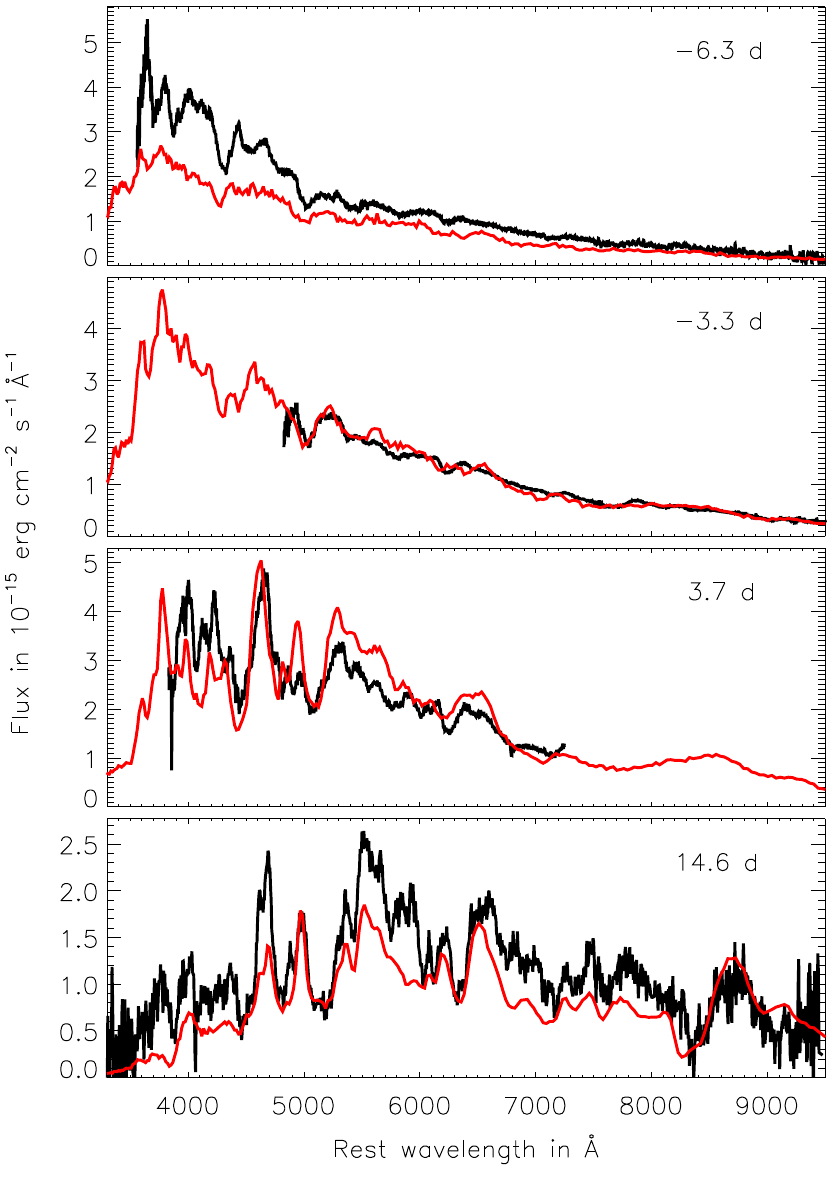}
  \caption{Spectral evolution of model N5def from -6.3 to 14.6 days
    (epochs are given with respect to $B$-band maximum).  For
    comparison the observed spectra of the 2002cx-like SN~2005hk
    \citep{phillips2007a} are over-plotted in black.  The observations
    were de-reddened and de-redshifted according to the values of
    \citet{phillips2007a} and \citet{sahu2008a}, respectively.}
  \label{fig:specplot}
\end{figure}

\citet{jha2006a} and \citet{phillips2007a} also report peculiar 
features for the spectra of 2002cx-like SNe.  At \emph{pre-maximum 
epochs} they find an \emph{SN~1991T-like spectrum} with a blue 
continuum, weak absorption features at ${\sim}\,4200$ and 5000\,\AA\
and no strong \ions{Ca}{ii} absorption features.  As can be seen 
from the top panels of Fig.~\ref{fig:specplot}, our model naturally 
predicts this behaviour, characteristic of an ionisation state
higher than in normal SNe~Ia.  

\begin{figure}
  \centering
  \includegraphics{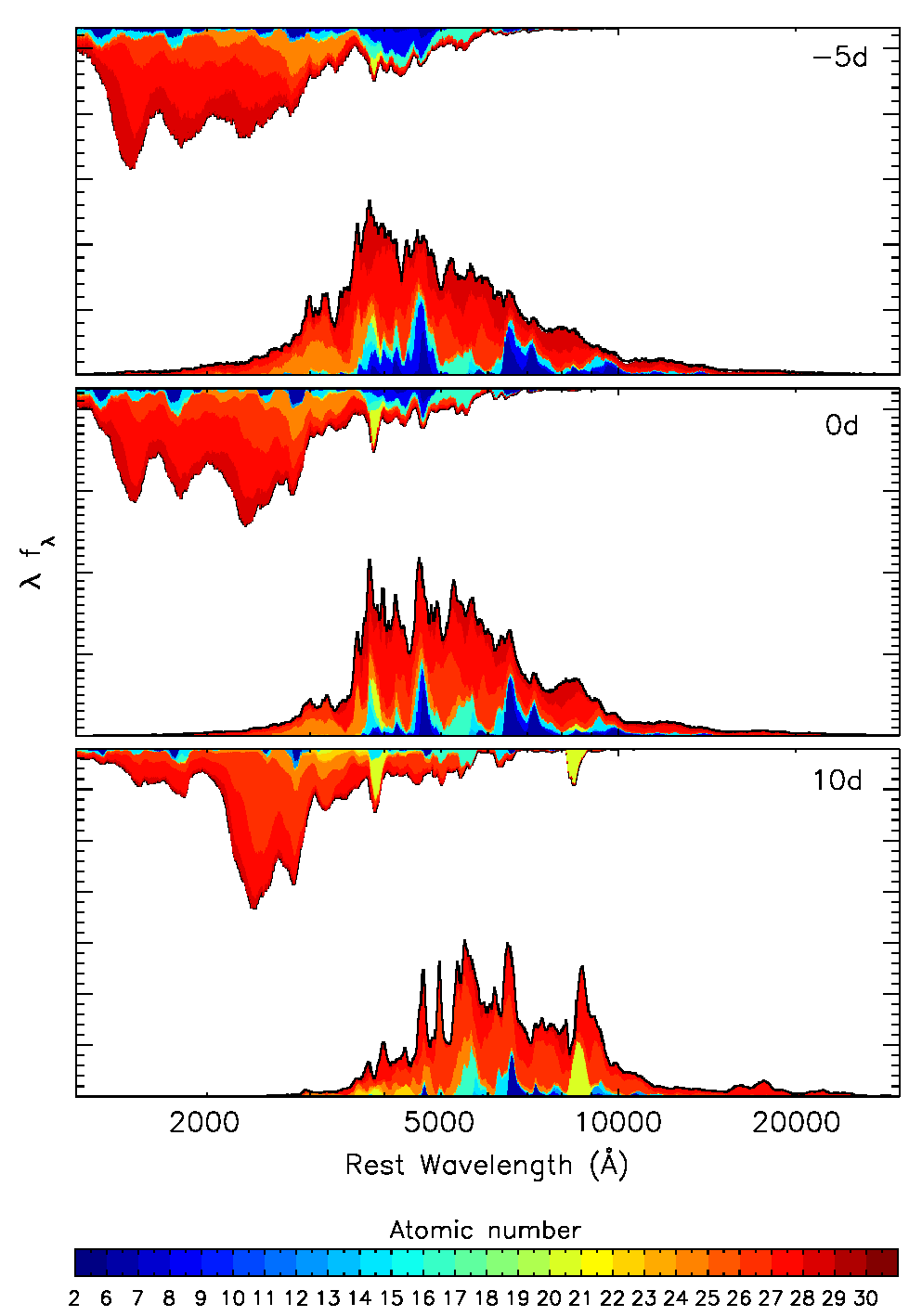}
  \caption{Formation of the UV to NIR SED for our N5def model for three
      different snapshots.  Epochs 
      are given relative to $B$-band maximum.  The colour coding 
      indicates the elements responsible for both bound--bound emission 
      and absorption of quanta in the Monte Carlo simulation.  The region below 
      the synthetic spectrum is colour-coded to indicate the fraction 
      of escaping quanta in each wavelength bin that were last emitted
      by a particular element (the associated atomic numbers are
      illustrated in the colour bar).  Similarly, the coloured regions
      along the top of the plots indicate which elements were last
      responsible for removing quanta from a particular wavelength bin
      (either by absorption or scattering\,/\,fluorescence).
      White regions between the emerging spectrum and the colour-coded 
      bound--bound emission indicate the contribution of continuum 
      processes (bound--free, free--free) to the last emission.}
  \label{fig:specelementsplot}
\end{figure}

Moreover, 2002cx-like SNe show a significant \emph{contribution 
from IGEs to the spectra 
at all epochs}, while normal SNe~Ia do not show strong IGE features 
until well-after maximum light.  Due to the strong mixing of the 
ejecta, our model is almost homogenized and thus predicts strong 
IGE contributions at all epochs.  
This is illustrated in Fig.~\ref{fig:specelementsplot}.  There, the
colour coding below the synthetic spectrum indicates the fraction of 
escaping packets in each wavelength bin that were last emitted by 
bound--bound transitions of a particular element (the associated 
atomic numbers are illustrated in the colour bar).  From this it is
clearly visible that emission by IGEs already dominates the spectrum 
before maximum light, while this is not the case for stratified
models, such as e.g. W7 (see figure~2 of \citealt{kromer2009a}).

Another characteristic feature of the spectra of 2002cx-like SNe
are \emph{extremely low line velocities at all epochs}, indicating
significantly lower expansion velocities and thus less kinetic
energy than in normal SNe~Ia.  Since our model burns only parts of 
the WD, the kinetic energy is rather low compared to other explosion 
models that fully burn and unbind the progenitor WD\@.  As a 
consequence, the line velocities of our model are in good agreement 
with the observed ones (compare Fig.~\ref{fig:specplot}).

\subsection{Near-infrared spectra}

\begin{figure}
  \centering
  \includegraphics[width=\linewidth]{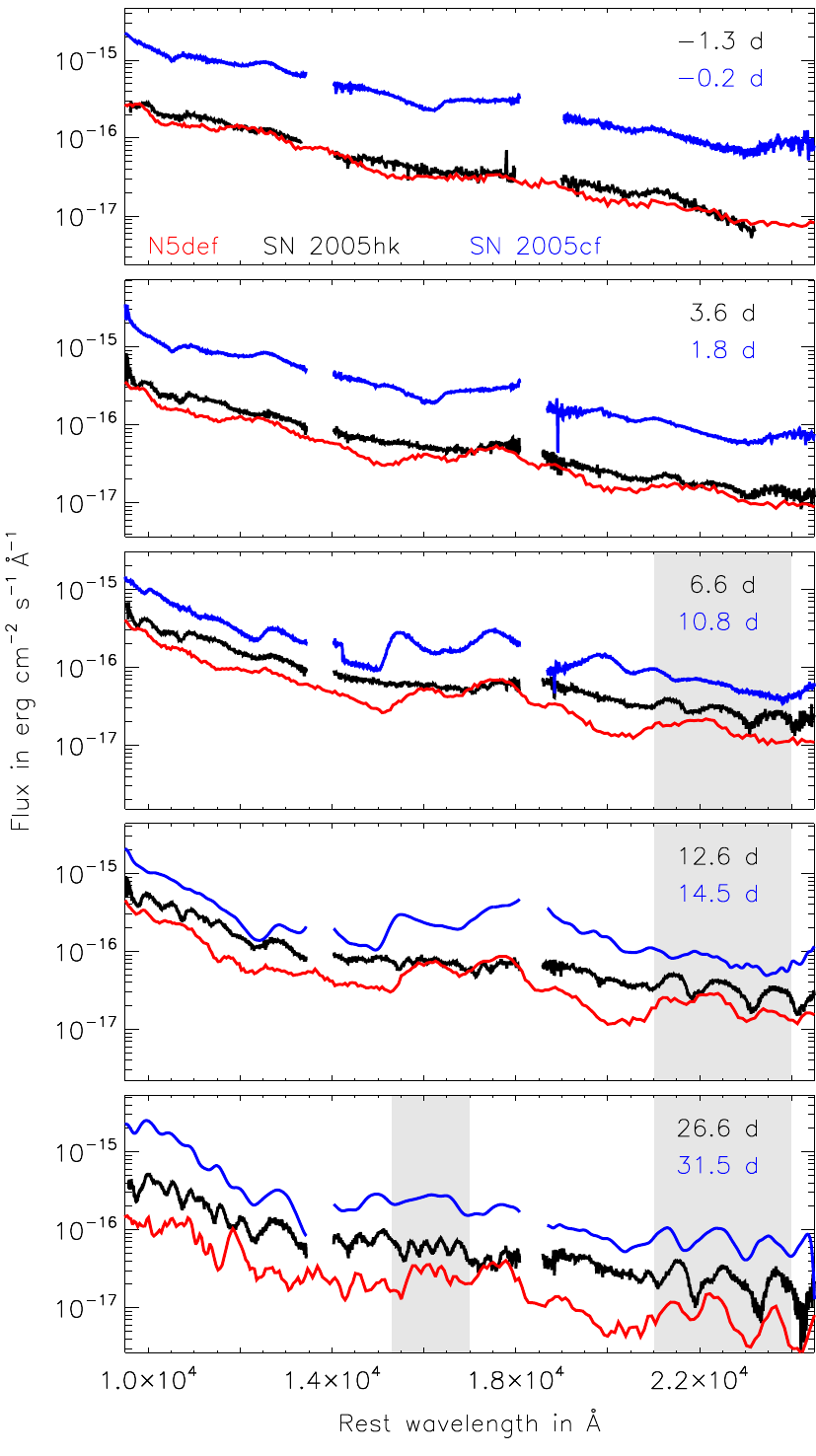}
  \caption{NIR spectra of the 2002cx-like SN~2005hk (black) and 
    synthetic spectra of model N5def (red) from $-1.3$ to 26.6 days
    (epochs are given with respect to $B$-band maximum).  The 
    observations were de-reddened and de-redshifted according to 
    the values of \citet{phillips2007a} and \citet{sahu2008a}, 
    respectively.  For comparison, NIR spectra of the normal SN~Ia
    2005cf \citep{gall2012a} are shown for similar epochs (blue).}
  \label{fig:nirspecplot}
\end{figure}

In Fig.~\ref{fig:nirspecplot} we show, for the first time, NIR 
spectra of a 2002cx-like SN\@.  These spectra of SN~2005hk were 
collected by the European
Research Training Network (RTN) `The Physics of Type Ia Supernova
Explosions'\footnote{http:/$\!$/www.mpa-garching.mpg.de/\url{~}rtn/}
(details on the observations and data reduction are given in 
Appendix~\ref{sec:nir_spectra_reduction}).  Compared to normal SNe~Ia 
there are a few prominent differences.  While in normal SNe~Ia the 
characteristic \ions{Co}{ii} emission humps in the $K$ band 
\citep[e.g.][]{gall2012a} do not form before 20\,d after maximum 
light, they emerge at 6.6\,d after maximum in SN~2005hk
(see grey highlighted region in Fig.~\ref{fig:nirspecplot}).  At later 
epochs ($\sim$\,30\,d) the \ions{Co}{ii} features appear significantly 
more narrow in SN~2005hk than in normal SNe~Ia.  A similar but even 
stronger effect can be observed in the $J$ and $H$ bands, where very
narrow spectral features develop at these epochs.  

Our model naturally reproduces these characteristic features of 
the NIR spectra of SN~2005hk.  The narrow features at later epochs
are a direct consequence of the low kinetic energy 
released during the explosion and reflect the low expansion
velocities of the ejecta (Fig.~\ref{fig:composition}) compared to 
that of models of normal SNe~Ia (e.g. \citealt*{nomoto1984a}, 
\citealt{roepke2012a}).  The early occurrence of the \ions{Co}{ii} 
emission humps can be understood as a result of the almost completely 
mixed ejecta.  While IGEs are concentrated in a central core in models 
of normal SNe~Ia, the turbulent burning of our deflagration model
leads to a very homogeneous IGE distribution in the ejecta, which 
extends to the highest velocities.
As shown by
\citet{kasen2006b}, such a mixing does not only lead to single 
peaked NIR light curves but also to an increased NIR flux around 
$B$-band maximum.  

Although our synthetic NIR spectra reproduce the characteristic
features of SN~2005hk, a detailed comparison shows some
differences.  While the absolute flux level and the overall
SED agree quite well with the observations around maximum
light, differences start to emerge about a week after maximum
light.  At these epochs the flux level of our model is too
low compared to the observations as already discussed for
the broad-band light curves (Section~\ref{sec:lightcurves}).

\section{Discussion}
\label{sec:discussion}

\subsection{Spectropolarimetry}
There are a few observational constraints which we cannot address
directly with our synthetic observables. \citet{chornock2006a} 
and \citet{maund2010a} presented spectropolarimetric observations 
of SN~2005hk and reported a line and continuum polarization on a 
level of a few tenths of a percent, which is typical of SNe~Ia 
\citep{wang2008a} and requires approximately spherical explosion 
ejecta.  Since our radiative transfer simulations currently do not 
include the polarization state of the radiation field, we 
cannot directly compare the observed polarization spectra to 
our model. We note, however, that our ejecta structure does not show 
very pronounced large-scale asymmetries (see Fig.~\ref{fig:composition}) 
and that the synthetic observables are not very sensitive to different 
lines-of-sight as visible from the light curves in 
Fig.~\ref{fig:lightcurves}. Thus, a weak polarization signal 
seems plausible.

\subsection{Late-time spectra}
\label{sec:late-time}
Further observational constraints come from late-time ($\sim$\,9\, 
months after explosion) spectra, where 2002cx-like SNe show extremely
narrow \emph{permitted \ions{Fe}{ii} lines and a continuum or 
pseudo-continuum like SED} \citep{jha2006a, phillips2007a}.  In 
contrast, normal SNe~Ia are in the nebular phase at these epochs.  
\citet{jha2006a} also report a tentative detection of \emph{weak 
\ions{O}{i} features} in the late-time spectra of SN~2002cx.
They do, however, not see strong features of [\ions{O}{i}] 
$\lambda\lambda$6300,6364 which have been predicted for 3D 
deflagration models \citep{kozma2005a}, where unburned O is present 
down to the lowest ejecta velocities due to turbulent burning
as in our model.

Since \textsc{artis} does not yet account for non-thermal ionization 
and excitation, which are the dominant processes at late epochs 
\citep[e.g.][]{kozma1998a, kozma1998b}, we cannot directly test the 
late-time spectra of our models against the observations. Whether 
[\ions{O}{i}] features will arise, however, depends strongly on a 
possible microscopic mixing of different species.   This cannot be 
addressed with present-day numerical models, since it takes place 
on scales which are not resolved.  However, if such a mixing is 
present, transitions of other species stronger than 
[\ions{O}{i}] $\lambda\lambda$6300,6364 will dominate the cooling.
Also the central density of our explosion ejecta ($\sim$\,120\,\gccm\,
at  100\,s after explosion) is about 30 times larger than in the 
W7 model ($\sim$\,4\,\gccm,\ \citealt{nomoto1984a}) or typical 
delayed detonation models that are in good agreement with normal 
SNe~Ia.  This might be important in explaining the peculiar features 
of the late-time spectra of 2002cx-like SNe.

Beyond that, we can only speculate that the bound remnant in our 
model might be able to explain the peculiar non-nebular features 
of the late-time spectra of 2002cx-like SNe if the observed 
spectrum was a superposition of emission from the bound
remnant and the SN ejecta.  We will discuss the implications of
the bound remnant in more detail in the next section.  Alternatively,
\citet{sahu2008a} tried to explain the observed late-time spectra of 
SN~2005hk by combining synthetic photospheric \citep[e.g.][]{mazzali1993a} 
and nebular \citep[e.g.][]{mazzali2001b} spectra of the ejecta of a 
parametrized 1D explosion model.

\subsection{Implications of the bound remnant}
\label{sec:remnant}

As discussed in Section~\ref{sec:explosion}, our model does not burn
the complete WD but leaves behind a bound remnant.  Due to the 
strong expansion of the ejecta, at late times this remnant cannot 
be resolved from our simulations, thus preventing any detailed analysis.  
However, it is clear that the remnant must be a puffed-up 
stellar object heated during the explosion.  From the bound tracer 
particles we can derive the overall composition of the remnant 
(see Table~\ref{tab:yields}), which shows that it is enriched with
IGE material and some amount of radioactive \nuc{56}{Ni}.  This 
enrichment is due to fall-back of explosion ashes onto
the bound remnant, so that the enriched material should be located
in the outer layers of the remnant (see also \citealt{jordan2012b}).

\begin{figure}
  \centering
  \includegraphics[width=\linewidth]{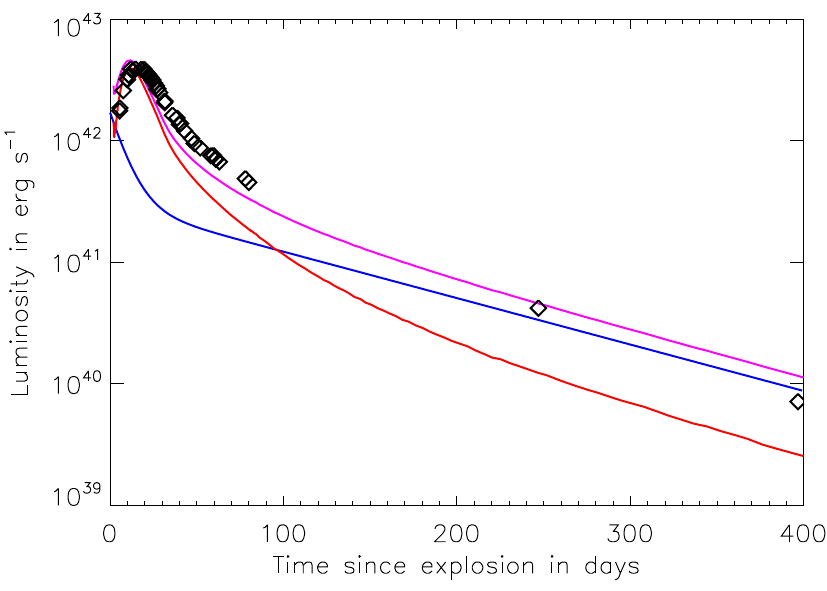}
  \caption{Bolometric light curve of SN~2005hk (black diamonds;
    following \citet{phillips2007a}, we adopted a rise time of 15\,d).
    For comparison, the red line shows the synthetic bolometric light 
    curve from our radiative transfer simulation in the ejecta. At
    $t=120$\,d our detailed non-grey radiative transfer simulation 
    ends.  Thereafter, we extended the synthetic light curve with a
    calculation using a grey UVOIR opacity.  The blue curve shows 
    the instantaneous energy deposition due to the \nuc{56}{Ni} 
    decay sequence in the bound remnant, assuming full $\gamma$-ray
    trapping.  The magenta curve shows the sum of both these
    contributions.}
  \label{fig:late-lc}
\end{figure}

We also note that the \nuc{56}{Ni} content of the bound remnant is 
large enough to contribute to the optical display of SN~2005hk at late
epochs (Fig.~\ref{fig:late-lc}) if $\gamma$-rays are completely
trapped in the remnant.  Since the remnant does not expand, its
densities are significantly larger than those of the ejecta,
making this assumption plausible.  Having some contribution
from the bound remnant at late epochs, could also explain the highly
peculiar spectra of 2002cx-like SNe, which never become fully nebular
(see discussion in previous section).  In particular the
extremely narrow IGE features might be explained if the emission
originated from an IGE-rich crust of the bound remnant.

Being heated by the actual explosion and long-lived radioactive 
isotopes, the bound remnant might even become directly detectable 
at much later epochs when the explosion ejecta go into the supernova 
remnant phase.  For SN~2005hk, observations were performed with the 
Gran Telescopio Canarias on July 20, 2012, i.e.\ 2443.5\,d after 
$B$-band maximum. These, however, have not shown any point source 
down to a limiting magnitude of $M_r = -8.08 \pm 0.49$ (for 
details of the observation see Appendix~\ref{sec:GTC_observation}).  

After the bound remnant has radiated away all the heat from the 
explosion and possible reheating from radioactive isotopes, it will
again become a WD, however with a peculiar composition enriched in
IGEs (and also intermediate-mass elements).  
Moreover, \citet{jordan2012b} report kick velocities of up to 
520\,\kms\ for the remnant WDs in their failed gravitationally-confined 
detonation simulations, thus 
claiming the existence of hyper-velocity enriched WDs as a `smoking
gun' for this scenario.  From our simulations we do not see such
kicks.  This difference may originate from the different gravity 
solvers used.  In general, no perfect momentum conservation is to
be expected from approximate solutions of the Poisson equation.

\subsection{Implications on the binary companion star}
\label{sec:companion}

One problem of single-degenerate progenitor models discussed
in the literature is that of the non-detection of hydrogen lines in
the observed (nebular) spectra of SNe~Ia \citep[e.g.][]{leonard2007a}.
Recent hydrodynamical simulations investigating the impact of SN~Ia 
ejecta on a non-degenerate companion star (e.g.\ \citealt*{pan2012a};
\citealt{liu2012a}) have shown that the amount of hydrogen stripped from the
companion star seems to be significantly larger than the observationally
derived upper limits \citep{leonard2007a}. Due to the low kinetic
energy of our explosion model, compared to that of normal SNe~Ia, only 
a small amount of material should be stripped from the companion
in the progenitor systems of SN~2002cx-like SNe, thus probably avoiding 
a signature of hydrogen lines in late-time spectra. This should be
addressed by detailed impact studies in the future.

\subsection{The class of 2002cx-like SNe}
\label{sec:02cx-class}

After the discovery of SN~2002cx by \citet{li2003a} as a peculiar
object, it soon became evident that this was not a unique event but
rather the prototype of a new class of peculiar SNe~Ia with a
striking spectral homogeneity \citep{jha2006a}.  From a volume-limited 
sample of the Lick Observatory Supernova Search (LOSS), \citet{li2011a}
estimate that SN~2002cx-like explosions contribute at about 
5 per cent to the total SN~Ia rate.

Recently, SN~2008ha \citep{foley2009a} and SN~2007qd 
\citep{mcclelland2010a} were proposed as additional members of the 
class. Though spectroscopically similar to SN~2002cx, those objects 
are fainter by more than 2 magnitudes than a typical 2002cx-like SN 
and show far lower line velocities, on the order of 2000\,\kms,\ while 
SNe~2002cx and 2005hk showed line velocities on the order of 
7000\,\kms.  Nevertheless, \cite{mcclelland2010a} found a 
relationship between light-curve stretch, the peak brightness 
and the expansion velocities among their (small) sample of objects.
This indicates that their extended class of SN~2002x-like objects 
may originate from a single explosion mechanism.

With an observationally derived \nuc{56}{Ni} mass of 0.003\,\msun\
for SN~2008ha \citep{foley2009a}, however, this object cannot
easily be explained in our model of a deflagration of a 
Chandrasekhar-mass WD\@.  From a systematic study of 3D full-star pure 
deflagration simulations for different ignition setups (Fink
et al., in preparation) we obtain a minimum \nuc{56}{Ni} mass
of 0.035\,\msun, which is more than a factor 10 larger than the
observationally derived value for SN~2008ha.  Though we did not
sample all possible ignition setups in the Fink et al.\ study 
nor take into account different progenitor compositions, this
difference may be difficult to reconcile with a deflagration 
of a Chandrasekhar-mass WD\@.  Thus another explosion mechanism
might be at work for these faint objects.  \citet{foley2009a,foley2010a} 
suggested deflagrations of sub-Chandrasekhar-mass WDs, while 
\citet{valenti2009a} and \citet{moriya2010a} favoured a 
core-collapse origin.

\subsection{Rates}

As mentioned in the previous section, \citet{li2011a} estimate from 
the LOSS that 2002cx-like SNe contribute at about 5 per cent to 
the total SN~Ia rate.  This is remarkably close to some theoretical
predictions for the contribution of single-degenerate 
hydrogen-accreting systems to the total SN~Ia rate.  From binary
population synthesis models, e.g.\ \citet*{ruiter2009a} find a 
value of $\sim$\,3 per cent (their model 1 with $\alpha_\mathrm{CE} 
\times \lambda=1$).  At the same time their predicted total SN~Ia 
rate for all progenitor channels considered (single-degenerate 
Chandrasekhar-mass scenario and double-degenerate mergers with a total
mass above the Chandrasekhar-limit) is comparable to the Galactic 
SN~Ia rate \citep[][figure~2]{ruiter2009a}.  Taking this at face 
value, it could be that all single-degenerate systems will end up 
as 2002cx-like SNe.  Normal SNe~Ia would then originate from other
progenitors like double-degenerate mergers \citep[e.g.][]{pakmor2012a, 
ruiter2012a} or sub-Chandrasekhar-mass double detonations 
\citep[e.g.][]{fink2010a, kromer2010a,woosley2011b}.  However, the 
observationally derived SN rates as well as the binary population 
synthesis rate predictions for the single-degenerate scenario are 
somewhat uncertain.  For example, \citet{han2004a} can reconcile the 
observed Galactic SN~Ia rate with the birth rate of SNe~Ia in the 
single-degenerate Chandrasekhar-mass scenario from calculations 
with their binary population synthesis code which assumes a 
different prescription for hydrogen accretion than \citet{ruiter2009a}.

\subsection{The fate of Chandrasekhar-mass explosions}

In a companion study (Fink et al., in prep.) we are currently 
investigating the outcome of a larger set of 3D deflagration
models. Within this sample we find a wide range of explosion 
strengths depending on the actual ignition conditions. However, 
to obtain \nuc{56}{Ni} masses on the order derived for SN~2002cx 
or SN~2005hk, only a very narrow range of ignition configurations
works.

We find that appropriate deflagrations with moderate
\nuc{56}{Ni} production result from ignition setups that can be 
interpreted in the scheme of a slightly off-centre single-spot 
ignition.  Thus, if all Chandrasekhar-mass explosions led to 
2002cx-like SNe, as hypothesized in the previous section, only 
such ignition configurations should be realized in Nature. Note 
that this is in agreement with state-of-the-art pre-ignition 
simulations \citep[e.g.][]{kuhlen2006a,zingale2009a,nonaka2012a}.

However, it could also be that for other configurations an 
ensuing detonation after the initial deflagration phase occurs, 
thereby producing a layered ejecta structure as needed for normal 
SNe~Ia \citep[e.g.][]{mazzali2007a}.  Several mechanisms to trigger 
such a secondary detonation have been proposed: a spontaneous 
deflagration-to-detonation transition \citep{khokhlov1991a},
a gravitationally confined detonation \citep{plewa2004a}, a pulsating 
reverse detonation \citep{bravo2006a} or a pulsationally-assisted 
gravitationally-confined detonation \citep{jordan2012a}.  However, 
for none of these models the formation of the secondary detonation 
can be resolved in detail.  Instead, detonations are triggered when 
certain conditions are met which are thought to be sufficient to 
ignite a detonation from simplified 1D simulations.

\section{Conclusion}
\label{sec:conclusion}

We have presented hydrodynamic simulations and radiative-transfer 
calculations for a 3D pure-deflagration of a Chandrasekhar-mass WD\@.
Our particular model (N5def), chosen from a larger sample of 
simulations (Fink et al., in preparation), has an asymmetric ignition 
configuration leading to a relatively weak deflagration, which fails 
to completely unbind the WD\@.  Only $\sim$\,0.37\,\msun\ of material 
are ejected, whereas the remainder stays gravitationally bound, 
leaving behind a stellar remnant of $\sim$\,1\,\msun.

The ejecta contain a significant \nuc{56}{Ni} mass ($\sim$\,0.16\,\msun),
have low kinetic energy, are well-mixed and relatively symmetric on 
large scales.  All of these properties make 
N5def a promising model for 2002cx-like SNe~Ia.  Moreover, 
a comparison of our synthetic observables with SN~2005hk shows good
agreement.  Our synthetic light curves reproduce the peak luminosity 
and colour of SN~2005hk and naturally predict the lack of a secondary 
maximum in the NIR bands.  Our synthetic spectra are also in good 
agreement with observations.  At optical wavelengths, N5def 
shows the same high ionisation, low expansion velocities and early 
dominance of IGE lines observed in SN~2005hk.  In addition, we have 
presented and compared with previously unpublished NIR spectra of 
SN~2005hk, allowing us to verify that the good match between model 
and observations extends into the NIR regime.
The main shortcoming of the model is an overly rapid post-maximum
fading at red and NIR wavelengths that is also apparent in the 
bolometric light curve, which is too narrow. This points at too low 
opacities, which could be the result of insufficient ejected mass.

The most intriguing feature of the model presented here is the bound 
remnant.  Unfortunately, at late times the remnant is not resolved in 
the simulation, so that its exact state after the explosion remains 
largely unknown.  We do find, however, that some fallback of 
nucleosynthetically processed material (including $\sim$\,0.02\,\msun\ 
of \nuc{56}{Ni}) onto the remnant will occur, resulting in a peculiar 
IGE-rich composition of the remnant object.  Heated by the explosion,
this will be a puffed-up stellar object.  We can also speculate that the 
bound remnant may be sufficiently luminous to contribute to the 
observed display of the SN, explaining in part the highly peculiar 
late-time spectra of 2002cx-like SNe, which do not become fully nebular 
even one year after the explosion.

While N5def seems to be a reasonable model for 2002cx-like SNe with 
\nuc{56}{Ni} masses of $\sim$\,0.2\,\msun, it appears unlikely to 
produce \nuc{56}{Ni} masses two 
orders of magnitude lower in the deflagration of a Chandrasekhar-mass 
WD\@.  These would be needed to explain 2008ha-like SNe, which 
have been suggested to be fainter and less energetic siblings 
of 2002cx-like SNe, originating from the same explosion mechanism.

Finally, we point out that the observed rate of 2002cx-like SNe 
(about 5 per cent of the total SN~Ia rate) is quite similar to the 
contribution of the single-degenerate Chandrasekhar-mass explosion 
channel to the total SN~Ia rate as expected from some binary population 
synthesis calculations.  It is thus tempting to speculate that 
deflagration-to-detonation transitions might actually not happen in 
SNe~Ia, and that single-degenerate Chandrasekhar-mass explosions 
might only lead to pure deflagrations with a 2002cx-like optical 
display.

\appendix
\label{sec:appendix}

\section{Observation log and data reduction of near-infrared spectra}
\label{sec:nir_spectra_reduction}

Five low-resolution NIR spectra of SN\,2005hk were obtained
at VLT/ISAAC (Table\,\ref{t:speclog:nir}).  During each night when the
SN was observed, one of the following bright stars Hip\,1115,
Hip\,26093 or Hip\,115347, all of spectral type B3V,
were also observed close in time and airmass to the SN observations.
Four different instrument settings were used to obtain wavelength
coverage between 0.97\,\mm\ and 2.5\,\mm. The observations for each
setting were reduced separately.  The spectra from the different
instrument settings that did not overlap were scaled so that the
synthetic $J$ and $H$ photometry matched
the observed one published by \citet{phillips2007a}.

The observations were performed in ABBA sequences, where A and B
denote two different positions along the slit. The images were
bias and flat field corrected, and trimmed. Because of the field
distortion of ISAAC, spectra taken at two different slit positions
are not parallel. Besides, the image of the slit is also curved. For the
subsequent data reduction it was necessary to correct for these two
effects. This was achieved with the TWODSPEC.LONGSLIT package in
IRAF\@. The input data for deriving this correction were: (i) the
arc-lamp spectra obtained during the nights when the SN was observed
and (ii) special (STARTRACE) observations obtained by ESO twice per
year and consisting of spectra of a bright star taken at 11 different
positions along the slit.  After applying the geometric distortion
correction the night sky lines were parallel to the image columns and
the spectra to the image rows.  Next, for each pair of AB images the B
image was subtracted from the A image. The negative spectrum was
shifted to the position of the positive one and subtracted from it,
resulting in an image with the sum of the spectra but without the sky
background. All such images were summed into a single image and the 1D
spectra were then optimally extracted.
We note that the optimal extraction algorithm has to be applied on
images with the pixel levels in the form of actual detected counts,
and so it does not work quite correctly if applied to
background-subtracted images. Special care was thus taken to calculate
the optimal extraction weights correctly.  The wavelength calibration
was done with arc-lamp spectra. Its accuracy was checked against the
sky lines and small corrections were applied if necessary.

To remove the strong telluric absorption features, the SN spectra were
divided by the high signal-to-noise spectra of the B3V stars. To
obtain the relative flux calibration, the result was multiplied by a
model continuum spectrum for a B3V star. We used a model spectrum with
$T_{\rm eff}=18500$\,K and $\log g=4$ by R.~Kurucz, available at
http://kurucz.harvard.edu/.  The residual features due to the
hydrogen absorption lines of the B3V stars were removed by dividing
the SN spectra by the composite spectrum for the B3V spectral type
from \citet{pickles1998a}.  The composite B3V spectrum was first normalized
to the continuum and smoothed to the instrumental resolution of each
of the four instrumental setups used.

\begin{table}
  \caption{Log of the VLT/ISAAC NIR spectroscopy}%
  \label{t:speclog:nir}
  \centering{%
  \begin{tabular}{@{}lcrcc@{}}
  \hline
  Date  & JD & Phase \\
  (UT) &   & (day)  \\
  \hline
  2005/11/09 & 2\,453\,683.58  &  $-$1.4   \\
  2005/11/14 & 2\,453\,688.58  &     3.6  \\
  2005/11/17 & 2\,453\,691.58  &  6.6      \\
  2005/11/23 & 2\,453\,697.58  &  12.6     \\
  2005/12/07 & 2\,453\,711.58  &  26.6    \\
  \hline
  \end{tabular}}
\end{table}

\section{GTC late-phase photometry}
\label{sec:GTC_observation}

Deep $r$-band \citep{fukugita1996a} images of the SN~2005hk 
explosion site with exposure times of $3 \times 200$\,s were 
obtained at the Gran Telescopio Canarias (GTC) equipped with 
OSIRIS on {\sc UT} 2012 July 20, 2443.5\,d after $B$-band maximum. 
The images were debiased, flat-field corrected, astrometrically 
aligned and averaged. An image quality of 0.9 arcsec 
was measured from the full width at half maximum of 
isolated stars.
The photometric zero point of the image was determined by 
comparing instrumental magnitudes of stellar objects with 
the apparent $r$-band magnitudes reported in the SDSS data 
release 6 catalogue \citep{adelman-mccarthy2008a}. 
Since no source was visible at the position of SN~2005hk, we 
estimated the limiting magnitude in our image by measuring 
faint stellar sources in similar locations as SN~2005hk. The 
apparent magnitude of the faintest of these sources, 
$r = 25.68 \pm 0.41$, was then adopted as limiting magnitude 
for the remnant of SN~2005hk. Note that this is $\sim$\,3.9 mag 
fainter than the last (Bessell) $R$-band magnitude of SN~2005hk, 
taken 381.6\,d after $B$-band maximum \citep{sahu2008a}.
With a colour excess of $E(B-V)_\mathrm{tot}=0.112$ mag and a 
distance modulus $\mu = 33.46 \pm 0.27$ mag, our limiting 
apparent magnitude translates into a limiting absolute magnitude 
of $M_r = -8.08 \pm 0.49$.

\section*{Acknowledgements}

This work was supported by the European Union's Human Potential
Programme `The Physics of Type Ia Supernovae' under contract
HPRN-CT-2002-00303, the Deutsche Forschungs\-gemeinschaft via
the Transregional Collaborative Research Center TRR 33 `The Dark
Universe', the Excellence Cluster EXC153 `Origin and Structure of
the Universe' and the Emmy Noether Program (RO 3676/1-1).
MK, MF, FKR and SAS also acknowledge financial support by
the Group of Eight/Deutscher Akademischer Austausch Dienst (Go8/DAAD)
exchange program. FKR is supported by the ARCHES prize of the 
German Ministry of Education and Research (BMBF).  

The simulations presented in this work were carried out at the John 
von Neumann Institute for Computing in J\"{u}lich (Germany) as part 
of projects hmu20 and pra026 within the Partnership for Advanced 
Computing in Europe (PRACE).

This work makes use of data collected at the 8.2-m Very Large 
Telescope (Cerro Paranal, Chile; program ID 076.D-0183) and the 
10.4-m Gran Telescopio Canarias (La Palma, Spain; program ID GTC19-12A).

\bibliographystyle{mn2e}

\label{lastpage}
\end{document}